\documentclass[epj,final]{svjour}
\usepackage{graphicx}
\usepackage{amsmath}
\usepackage{amssymb}

\renewcommand{\d}{\mathrm{d}}
\renewcommand{\c}{\mathrm{c}}
\newcommand{\p}{\mathrm{p}}
\renewcommand{\in}{\mathrm{in}}

\begin{document}

\title{Cold trapped atoms detected with evanescent waves}
\author{R.A. Cornelussen \and A.H. van Amerongen \and B.T. Wolschrijn \and R.J.C. Spreeuw \and H.B. van Linden van den Heuvell}
\institute{Van der Waals - Zeeman Institute, University of Amsterdam,
Valckenierstraat 65, 1018 XE  Amsterdam, The Netherlands
\email{ronaldc@science.uva.nl}}

\PACS{{32.80.Pj}{Optical cooling of atoms; trapping}\and{42.25.Bs}{Wave propagation, transmission and absorption}}

\abstract{We demonstrate the {\em in situ} detection of cold $^{87}$Rb atoms
near a dielectric surface using the absorption of a weak, resonant evanescent
wave. We have used this technique in time of flight experiments determining the
density of atoms falling on the surface. A quantitative understanding of the
measured curve was obtained using a detailed calculation of the evanescent
intensity distribution. We have also used it to detect atoms trapped near the
surface in a standing-wave optical dipole potential. This trap was loaded by
inelastic bouncing on a strong, repulsive evanescent potential. We estimate
that we trap $1.5 \times 10^4$ atoms at a density 100 times higher than the
falling atoms.}

\maketitle

\section{Introduction}

Recently there has been increased interest in cold atoms trapped near a
surface. For example magnetic fields of micro-electronic structures are used to
trap and guide atoms near the surface of so-called atom chips
\cite{HanReiHan01,MulCorAnd00,CasHesSch00}. Last year Bose-Einstein
condensation was realized on such a chip \cite{HanHomRei01,OttForZim01}. Other
examples are experiments aiming to increase the phase space density and create
a quasi 2D gas of atoms using inelastic reflections from an evanescent-wave
(EW) atomic mirror \cite{SprVoiHeu00,OvcManGri97,GauHarMly98}.

These experiments pose new challenges for in situ detection, in particular if
the atoms are within the order of an optical wavelength from the surface. In
this case absorption of EWs is advantageous, since only atoms that are within a
few wavelengths from the surface are probed. Aspect {\it et al.} have proposed
the nondestructive detection of atoms close to a surface by detecting a phase
change of a far detuned EW \cite{AspKaiWes95}. However this effect is very
small and has not been observed so far. In this letter we demonstrate
experimentally the absorption of resonant evanescent waves as a novel
diagnostic tool to study cold atoms near a dielectric surface. Using a weak,
resonant EW, we have studied a sample of cold atoms falling onto the surface as
well as atoms trapped close to the surface.

EW absorption has previously been used for spec\-tros\-copy on hot and dense
atomic vapors to experimentally investigate EW properties such as the
Goos-H\"{a}nchen shift \cite{KieJozDoh98,ZhaWuLai01}.

\section{Evanescent wave calculations}

\begin{figure}[t]
\centerline{\scalebox{.5}{\includegraphics{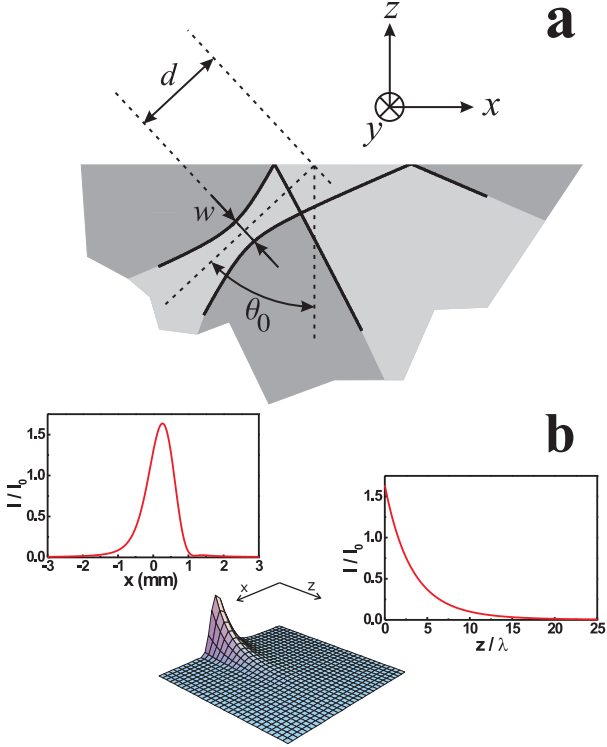}}} \caption{{\bf (a)}
Overview of the geometry and notations for the evanescent wave calculations
near the critical angle. The incident beam has a waist $w$ and angle of
incidence $\theta_0$. The waist is at a distance $d$ from the surface. {\bf
(b)} Intensity distribution of the evanescent wave for realistic experimental
parameters $w~=~330~\mu$m, $(\theta_0~-~\theta_\c)~=~133~\mu$rad and
$d~=~680~$mm. Transverse $x$-distribution at the prism surface,
$z$-distribution at the $x$-coordinate where the intensity at the surface is
highest.} \label{fig1}
\end{figure}

An evanescent wave appears when a light wave undergoes total internal
reflection at the surface of a dielectric with index of refraction $n$. For a
plane incident wave the optical field on the vacuum side of the surface decays
exponentially $\sim\exp(-z/\xi)$ with $z$ the direction perpendicular to the
surface, $\xi(\theta)=\frac{\lambda}{2\pi}(n^2\sin^2\theta-1)^{-1/2}$ the decay
length, $n$ the index of refraction of the substrate and $\theta$ the angle of
incidence. For a low density of resonant absorbers near the surface, the
scattering rate in the low saturation limit is proportional to the square of
the field: $\sim\exp(-2 z/\xi)$. If the density of absorbers is uniform, this
gives rise to a total rate of scattered photons proportional to $\xi$. The
scattered photons are missing from the reflected beam, which is therefore
attenuated. If the angle of incidence approaches the critical angle
$\theta_\c=\arcsin(n^{-1})$, the value of $\xi$ diverges, so the absorption
becomes large. The absorption is however less height selective in this limit.

For a Gaussian beam with an angle of incidence $\theta_0\gg\theta_\c+\phi$ with
$\theta_\c$ the critical angle and $\phi$ the divergence of the beam, the
electric field is similar to the field of a plane wave. For Gaussian beams with
$\theta_0$ closer to $\theta_\c$, the evanescent field does not decay as a
simple exponential. We can describe the incident field as a superposition of
plane wave Fourier components with various $\theta$. Each component contributes
an evanescent field component with decay length $\xi(\theta)$ and an amplitude
proportional to the complex Fresnel transmission coefficient $t(\theta)$.
Because both $\xi(\theta)$ and $t(\theta)$ vary strongly around $\theta_\c$,
the evanescent wave contributions of these incident plane wave components add
up to a non exponentially decaying field. In addition the transverse beam
profile is modified to a non Gaussian shape. In the reflected beam one of the
effects is the well known Goos-H\"anchen shift \cite{GooHan47}. Other phenomena
like nonspecular reflection, shift of focus point and change of beam waist have
been predicted \cite{ChiTam87}. They all result from combining a finite size
beam with angle dependence of the transmission coefficient $t(\theta)$.
Recently, it has been proposed to measure a Goos-H\"anchen shift also in the
evanescent wave using a scanning tunnelling optical microscope (STOM)
\cite{BaiLabVig00}.

In the following calculations we will coherently add plane wave components with
propagation vectors in the plane of incidence. The transverse distribution has
been assumed Gaussian with the same radius as the incident beam at the surface.
This approach is valid, since the transverse components of the propagation
vector change the angle of incidence only in second order.

The evanescent field has been calculated by evaluating the following expression
\begin{eqnarray}
E(x,z) = \frac{1}{\sqrt{\pi} \phi}\int\limits_{\theta_\c}^{\pi/2}
&t_{s,p}(\theta) p(\theta) \exp[i k_x(\theta) x - \frac{z}{\xi(\theta)}
\nonumber \\ &+ i n k_0 \frac{d}{2} (\theta-\theta_0)^2 -
\frac{(\theta-\theta_0)^2}{\phi^2} ] \d\theta, \label{eq1}
\end{eqnarray}
with $k_x(\theta)=n k_0 \sin\theta$ the wavevector along the prism surface,
$k_0=2\pi/\lambda$ and $\theta_0$ is the angle of incidence. The first two
exponents are the field distributions parallel and perpendicular to the
surface, respectively. The third exponent takes into account that the waist of
the incident beam is at a distance $d$ from the surface. The fourth exponent is
the distribution over angles of incidence for a Gaussian beam, with $\phi=2/(n
k_0 w)$ the $1/e$ half width of the angle distribution of the field of a
Gaussian beam with waist $w$. The factor $t_{s,p}(\theta)$ is the Fresnel
transmission coefficient for transmission of a $s$ or $p$ polarized plane wave
with angle of incidence $\theta$ from the dielectric into vacuum. They are
given by $t_s(\theta)=2 n \cos\theta/(n \cos\theta+i\sqrt{n^2\sin^2\theta-1})$
and $t_p(\theta)=2 n \cos\theta/(\cos\theta+i n \sqrt{n^2\sin^2\theta-1})$
respectively. Finally $p(\theta)$ is a normalization factor that is equal to 1
for $s$ polarized incident light and $\sqrt{2 n^2 \sin^2\theta-1}$ for $p$
polarized light. The integration is carried out over the range of angles of
incidence that generate an evanescent wave, from the critical angle $\theta_\c$
to $\pi/2$. The normalization is chosen such that $|E|^2=1$ in the waist of the
{\em incident} beam. The geometry of the problem and some of the parameters are
displayed in Fig. \ref{fig1}a. The effective evanescent intensity is given by
\begin{equation}
I(x,y,z)=\frac{1}{n} I_0 \frac{w}{w_\p} |E(x,z)|^2 e^{-2 y^2/w_\p^2},
\label{eq2}
\end{equation}
where $I_0$ is the intensity of the incident beam in the waist of the incident
beam. The Gaussian determines the distribution in the y-direction with $w_\p$
the transverse $1/e^2$ intensity radius at the prism surface. The fraction
$w/w_\p$ corrects for the transverse divergence of the incident beam. This
approach is possible since $|E|^2=1$ in the waist of the incident beam. Fig.
\ref{fig1}b shows a calculation of the EW intensity distribution for realistic
experimental parameters $(\theta_0-\theta_\c)~=~133~\mu$rad, $w~=~330~\mu$m and
$d~=~680~$mm. This distribution is used to calculate the absorption {\em via}
the integrated scattering rate.

\section{Time of flight experiments}

\begin{figure}[t]
\centerline{\scalebox{.5}{\includegraphics{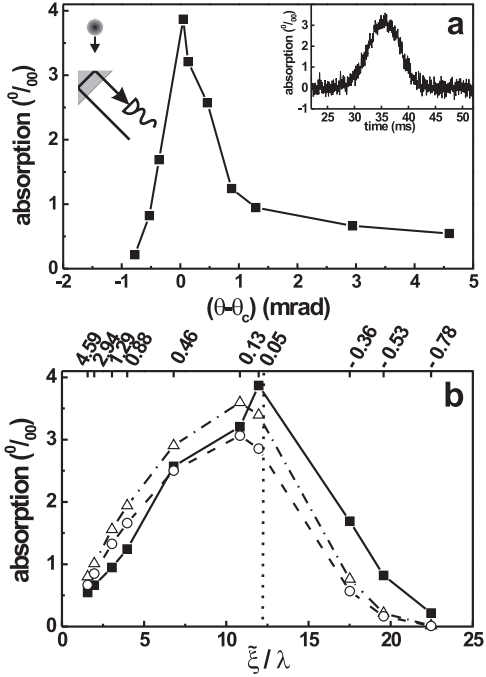}}} \caption{{\bf (a)}
measured absorption by a cloud of cold atoms falling onto a dielectric surface
as a function of the angle of incidence of the evanescent probe beam. The inset
shows a typical time of flight signal, measured with an angle of incidence of
$(\theta_0-\theta_\c)~=~130~\mu$rad. {\bf (b)} same data as shown in (a), but
{\em versus} effective decay length $\tilde{\xi}$. Also the results of the
calculations based on a density of $1.2\times10^9~$cm$^{-3}$ are shown.
$\blacksquare$ Measured data, $\bigcirc$ calculated absorption on the basis of
scattering, $\triangle$ calculations using complex index of refraction. The
vertical dotted line corresponds to the critical angle.} \label{fig2}
\end{figure}

In a first experiment we use evanescent wave absorption to detect a cloud of
cold atoms falling on a glass surface. Our setup consists of a right angle,
uncoated, BK7 prism, mounted inside a vacuum cell with a background Rb pressure
of $10^{-7}~-~10^{-8}~$mbar. About $2 \times 10^7$ $^{87}$Rb atoms are captured
from the background gas in a magneto optical trap (MOT), located 7~mm above the
prism surface and are subsequently cooled to 4.5$~\mu$K in optical molasses.
They are released in the F=2 hyperfine ground state and fall towards the prism
due to gravity. Inside the prism a weak, resonant, $p$-polarized probe beam
undergoes total internal reflection. Its angle of incidence was determined with
$100~\mu$rad accuracy by a method that will be discussed later. The probe beam
has a waist of $330~\mu$m ($1/e^2$ radius) resulting in a divergence of
500~$\mu$rad inside the prism. At the prism surface it has a waist ($1/e^2$
intensity radius) of $(470 \pm 20)~\mu$m. The total power of the incident beam
is $P_\in~=~2.2~\mu$W.

We measure the absorption in the reflected wave on a photodiode which is
amplified by a factor $10^6$ by a fast, low noise current amplifier (Femto
HCA-1M-1M). A typical absorption trace is shown in the inset of Fig.
\ref{fig1}a. The maximum absorption of time traces for different values of the
angle of incidence are plotted in Fig. \ref{fig2}a. From this graph it is clear
that the absorption is highest for angles of incidence very close to the
critical angle. In order to analyze these results we consider the atomic
density to be uniform perpendicular to the surface, since the penetration of
the EW ($<10~\mu$m) is much smaller than the size of the cloud ($\sim1~$mm). It
is crucial to take the finite probe beam diameter into account. This leads to a
finite range of angles of incidence so that the EW is no longer exponential as
described above. We define an effective decay length $\tilde{\xi}$ by
$\tilde{\xi}/2=\sqrt{\langle z^2 \rangle - \langle z \rangle^2}$ where the
distribution function is the normalized intensity distribution
$I(x_0,0,z)/\int_0^{\infty}I(x_0,0,z)\d z$ at the transverse position $x_0$
where the intensity at the surface is maximum. For a plane incident wave
$\tilde{\xi}=\xi$. In Fig. \ref{fig2}b the solid squares represent the same
absorption data as shown in Fig. \ref{fig2}a, but plotted {\em versus}
$\tilde{\xi}$. Absorption increases with $\tilde{\xi}$, but decreases beyond a
value $\tilde{\xi}\approx 12\lambda$. This decrease for larger $\tilde{\xi}$
occurs because the amplitude of the EW quickly drops for angles of incidence
$\theta_0$ smaller than $\theta_\c$.

We compare our data to the absorption as calculated using two different
approaches. The first method is to calculate the scattering of evanescent
photons by the atoms near the surface, where we assume that the transmission
coefficients are not changed by the atoms. The scattered power is calculated as
\begin{equation*}
\frac{1}{2}\hbar\omega\Gamma\int_{\mathrm{EW}} \rho(\vec{x})
\frac{s(\vec{x})}{s(\vec{x})+1} \d^3 x,
\end{equation*}
where $s(\vec{x})=\frac{7}{15}I(\vec{x})/I_{\mathrm{sat}}$ is the local
saturation parameter, $I_{\mathrm{sat}}$ is the saturation intensity
$1.6~$mW/cm$^2$, $I(\vec{x})$ is the local evanescent intensity, given by Eq.
(\ref{eq2}), $\rho(\vec{x})$ is the local density, $\hbar\omega$ is the energy
of a photon and $\Gamma~=~2\pi\times6~$MHz is the natural linewidth. The factor
$\frac{7}{15}$ appears because linearly polarized light is used. The
integration is over the entire volume of the evanescent wave. Because the
absorption is so low, Beer's law remains in its linear regime. Obviously
saturation effects are taken into account. We also account for the Van der
Waals potential, which leads to a decrease in the atomic density close to the
surface. Finally also the transverse density distribution of the atomic cloud
is taken into account. Neglecting these last three effects would increase the
calculated absorption by approximately 20\%. The open circles in Fig.
\ref{fig2}b are the results of this calculation for a density near the prism of
$1.2\times 10^9~$cm$^{-3}$.

For another measurement (not shown) with an angle of incidence
$(\theta_0-\theta_\c)~=~130~\mu$rad and an evanescent probe power of
$2.9~\mu$W, the measured maximum absorption of $(0.23\pm0.01)\%$ resulted in a
calculated density of $(1.3\pm0.4)\times 10^9~$cm$^{-3}$. This value agrees
very well with the density of $(1.3\pm0.2)\times 10^9~$cm$^{-3}$ derived from
conventional absorption images with a CCD camera.

Close to a dielectric surface the radiative properties of atoms are altered
\cite{CouCouMer96}. The natural linewidth of a two-level atom can be up to 2.3
times larger than the natural linewidth in free space. However, including this
effect in the calculations only increased the calculated absorption by about
2\%, which is well within the measurement uncertainty. By decreasing both the
probe intensity and the decay length this effect should start to play a role.

As a cross check, the second method determines the absorption by describing the
atomic cloud by a complex index of refraction $n=1+i \frac{\sigma_0 \rho
\lambda}{4 \pi}$, with $\sigma_0=\frac{7}{15}\frac{3\lambda^2}{2\pi}$ the
resonant cross section for linearly polarized light and $\rho$ the density at
the prism surface. Using this index of refraction to calculate the reflected
intensity also yields the absorption. The reflected field is determined by
evaluating
\begin{eqnarray}
E(x,z) = \frac{1}{\sqrt{\pi} \phi} \int\limits_{0}^{\pi/2} & r_p(\theta,n) \exp[i k_x(\theta) x - ik_z(\theta) z \nonumber \\
& + i n k_0 \frac{d}{2} (\theta-\theta_0)^2 -
\frac{(\theta-\theta_0)^2}{\phi^2}] \d\theta, \label{eq3}
\end{eqnarray}
with $k_z(\theta)=n k_0 \cos\theta$ the wavevector perpendicular to the surface
and
$r_p(\theta)=(\cos\theta-n\sqrt{1-n^2\sin^2\theta})/(\cos\theta+n\sqrt{1-n^2\sin^2\theta})$
is the Fresnel coefficient for reflection for $p$ polarized light. The same
normalization as for Eq. (\ref{eq1}) was used. The reflected intensity is given
by $I(x,z)=I_0 |E(x,z)|^2$. Saturation effects are not included. Since finally
only the total absorbed power is important, it is not necessary to incorporate
the transverse distribution in these calculations. The open triangles in Fig.
\ref{fig2}b show the results of these calculations for various angles of
incidence. The absorption for a maximum density near the prism surface of
$1.2\times 10^9~$cm$^{-3}$ calculated with the complex index of refraction is
slightly higher than the absorption calculated from the scattering of
evanescent photons, mainly because saturation effects were neglected.

\section{Trapping}

In a second experiment we used evanescent waves to detect atoms trapped close
to the surface in a standing light field. We load and trap the atoms using the
scheme as described in previous work \cite{SprVoiHeu00}. Cold atoms are
prepared as in the time of flight experiment. During their fall the atoms are
optically pumped to the F$=1$ hyperfine ground state. On the vacuum side of the
prism surface a repulsive EW potential is created by total internal reflection
of a 90~mW, TM polarized laser beam with a waist of 500$~\mu$m and blue detuned
by 0.2 - 1~GHz with respect to the F$=1~\leftrightarrow~$F'$=2$ transition of
the D1 line. This potential acts as a mirror for atoms. The decay length of the
EW can be controlled by changing the angle of incidence of the laser beam
\cite{VoiWolHeu00}. By scattering EW photons, the atoms can make a spontaneous
Raman transition to the F=2 hyperfine ground state, for which the repulsive EW
potential is lower. This results in a virtually complete loss of their
gravitationally acquired kinetic energy \cite{WolVoiHeu01}.

The trapping laser is a linearly polarized, 1.3W laser beam, red detuned by
about 85 GHz with respect to the D2 line. It is reflected from the vacuum side
of the uncoated prism surface, creating a standing wave (SW) with a visibility
of 0.38. The angle of incidence is nearly normal, 25~mrad. The spot size at the
prism surface is $380~\mu$m $\times$ $440~\mu$m ($1/e^2$ radii). For atoms in
the F=1 hyperfine ground state the EW potential is dominant, whereas for the
F=2 hyperfine ground state the SW potential is dominant. Atoms that are Raman
transferred to F=2 near their motional turning point can be trapped in the SW
potential. Only a few potential minima will be occupied due to the highly
localized optical scattering of the EW. When the atoms fall towards the prism
both the EW and the SW are off. Both are turned on 1~ms before the maximum atom
density reaches the prism surface. In order to decrease the scattering rate,
the EW is switched off after 2~ms, after which the loading of the SW trap
stops.

The EW probe beam is aligned by overlapping it with the EW bouncer beam, whose
angle can be set with an accuracy of 25~$\mu$rad. The overlap could be checked
over a distance of 1.5~m, resulting in an uncertainty in the angle of
$100~\mu$rad. During the trapping experiments the probe angle of incidence was
kept constant at $(\theta_0-\theta_\c)~=~130~\mu$rad and the power of the probe
beam is $P_\in~=~2.2~\mu$W. At the prism surface it had a waist ($1/e^2$
intensity radius) of $(770\pm10)~\mu$m. The probe was resonant for atoms in the
$F=2$ hyperfine ground state and was turned on at least 14~ms after shutting
the EW bouncer beam in order to be certain that no falling atoms were present
and only trapped atoms would be probed.

\begin{figure}[t]
\centerline{\scalebox{.5}{\includegraphics{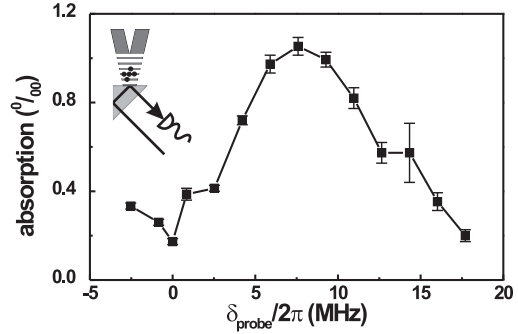}}} \caption{Measurement of
atoms trapped in a standing wave, detected using evanescent probing. The
absorption for different probe detunings shows that the atoms are distributed
over several light shifts. We determine that $1.5 \times 10^4$ atoms are
initially trapped.} \label{fig3}
\end{figure}

Since the trap is red detuned, the atoms will be trapped in the intensity
maxima. In the center of the trap the scattering rate in these maxima is
calculated to be 7~ms$^{-1}$ and the ground state light shift is
$\delta_{\mathrm{LS}}/2\pi~=~-15.4$~MHz. The trap depth here is only $8.6$~MHz
since the fringe visibility of the standing wave is 0.38. The trap frequency is
$359$~kHz, which results in approximately 24 bound levels. The resonance
frequency of trapped atoms is blue-shifted by $-2 \delta_{\mathrm{LS}}$, due to
the light shift of the excited state.

Fig. \ref{fig3} shows the absorption of an evanescent probe beam by atoms
trapped in the standing wave. The evanescent wave bouncer is blue detuned by
550~MHz and has a decay length $\xi=1.15\lambda$. The evanescent probing
technique was used to optimize the trap loading by varying the bouncer
detuning. The maximum absorption is observed for a detuning of the EW probe of
8~MHz. The measured linewidth is larger than the 6~MHz atomic linewidth,
probably due to inhomogeneous broadening. There are two contributions to the
broadening. Firstly, since the trap laser has a Gaussian profile the atoms
encounter a spatially varying light shift. Secondly atoms in higher excited
vibrational states will encounter a smaller light shift. It is not possible to
reliably retrieve these detailed distributions from our measured curve. It is
however possible to make a good estimate of the total number of trapped atoms.
The relative absorption curve in Fig. \ref{fig3} is, for low saturation,
described by
\begin{equation}
\frac{-\delta P}{P}~=~\frac{\hbar\omega\Gamma}{P_\in} \int\int
\frac{s(\vec{x})}{2}~\frac{\tilde{\rho}(\vec{x},\Delta)}{1+4\left(\frac{\delta-\Delta}{\Gamma}\right)^2}~\d^3\vec{x}~\d\Delta,
\label{eq4}
\end{equation}
which is similar to the scattering analysis of the falling atoms, but
inhomogeneously broadened by means of the convolution with the Lorentz
function. The factor $P_\in$ is the power of the incident probe beam and
$\tilde{\rho}(\vec{x},\Delta)$ is the distribution of atoms in the $F=2$
hyperfine ground state over spatial coordinates and light shifts. By
integrating Eq. (\ref{eq4}) over the detuning $\delta$ (the area under the
curve of Fig. \ref{fig3}), the integration of the Lorentzian over $\delta$
yields $\pi\Gamma/2$. The integration over $\Delta$, for which the integrand is
now only $\tilde{\rho}(\vec{x},\Delta)$, yields the density of atoms
$\rho(\vec{x})$.

From comparing the kinetic energy of the falling atoms in $F=1$ to the trap
depth when the atoms are pumped to $F=2$, it follows that mainly the third and
fourth potential minima will be populated. An underestimate of the number of
atoms can be obtained by assuming the probe intensity constant over the trap
region (which is valid if the evanescent probe size is much larger than the
trap size) and all atoms are in the third minimum. Eq. (\ref{eq4}) then reduces
to $(\pi\hbar\omega\Gamma^2 s_3)/4 P_\in)N$, with $N$ the total number of
trapped atoms and where $s_3$ denotes the saturation parameter in the third
potential minimum. From this a number of trapped atoms in the $F=2$ hyperfine
ground state of $3.0 \times 10^3$ can be calculated after 14~ms of trapping.
The total number of trapped atoms will be $4.5 \times 10^3$ because the steady
state $F=1$ groundstate population will be 0.5 times the $F=2$ population due
to scattering by the standing wave with the present detuning. The populations
of the magnetic sublevels of one hyperfine groundstate are equal to within 1\%.
By comparison with previous measurements we deduce a lifetime of 14.3~ms. An
extrapolation results in about $1.2 \times 10^4$ trapped atoms at $t=0$. The
assumption that the evanescent probe size is much larger than the trap size is
not completely valid. Taking the correct radii into account leads to a 22\%
increase, thus $1.5 \times 10^4$ trapped atoms. Assuming the transverse trap
distribution equal to the trapping laser, the vertical trap radius to be
$\lambda/4$ and the atoms to be distributed mainly over two potential minima,
the density becomes $1.2 \times 10^{11}~$cm$^{-3}$, which is about 100 times
higher than the density of the atoms falling onto the prism.

\section{Conclusions and outlook}

We have shown that the absorption of a weak, resonant evanescent wave can be
used to selectively detect cold atoms near ($\sim\lambda$) a dielectric
surface. A model treating the absorption by scattering evanescent photons was
suitable to describe the absorption. When calculating the evanescent intensity
distribution, the Gaussian character of the incident beam had to be taken into
account in order to quantitatively understand the absorption for angles of
incidence very close to the critical angle.

By detecting cold atoms falling onto the dielectric surface for different
angles of incidence of the probe beam we have verified our quantitative
description of the dependence of the absorption on the angle of incidence of
the probe beam. By detecting cold atoms trapped in standing wave potential
minima close to the surface we have determined that we have trapped more than
$1.5 \times 10^4$ atoms initially. This results in an increase of the density
of two orders of magnitude with respect to the approaching atoms.

The technique can be extended to using a CCD camera so that a transverse
distribution of the atoms can be measured. By performing measurements for
different angles of incidence of the probe beam, it should be possible to
obtain information about the distribution of atoms perpendicular to the
surface.

\begin{acknowledgement}
This work is part of the research program of the ``Stichting voor Fundamenteel
Onderzoek van de Materie'' (Foundation for the Fundamental Research on Matter)
and was made possible by financial support from the ``Nederlandse Organisatie
voor Wetenschappelijk Onderzoek'' (Netherlands Organization for the Advancement
of Research). R.S. has been financially supported by the Royal Netherlands
Academy of Arts and Sciences.
\end{acknowledgement}

\end{document}